\documentclass[aps,prd,groupedaddress, twocolumn, eqsecnum,nofootinbib,showpacs,preprintnumbers]{revtex4}
\usepackage{graphicx,epsf,amssymb,amsbsy,amsfonts,amssymb,amsmath}
\usepackage[usenames, dvipsnames]{color}
\usepackage{pst-node,pstricks-add}
\usepackage{calc}
\usepackage{url}
\newif\ifdraft
\drafttrue
\newif\ifpreprint
\preprinttrue

\def\sect#1{section~{\ref{#1}}}
\def\fig#1{fig.~{\ref{#1}}}

\def\f{\tilde f}

\def\Tr{\, {\rm Tr}}

\def\spa#1.#2{\left\langle#1\,#2\right\rangle}
\def\spb#1.#2{\left[#1\,#2\right]}
\def\spash#1.#2{\spa{\smash{#1}}.{\smash{#2}}}
\def\spbsh#1.#2{\spb{\smash{#1}}.{\smash{#2}}}
\def\sand#1.#2.#3{%
\left\langle\smash{#1}{\vphantom1}^{-}\right|{#2}%
\left|\smash{#3}{\vphantom1}^{-}\right\rangle}
\def\sandpp#1.#2.#3{%
\left\langle\smash{#1}{\vphantom1}^{+}\right|{#2}%
\left|\smash{#3}{\vphantom1}^{+}\right\rangle}
\def\sandpm#1.#2.#3{%
\left\langle\smash{#1}{\vphantom1}^{+}\right|{#2}%
\left|\smash{#3}{\vphantom1}^{-}\right\rangle}
\def\sandmp#1.#2.#3{%
\left\langle\smash{#1}{\vphantom1}^{-}\right|{#2}%
\left|\smash{#3}{\vphantom1}^{+}\right\rangle}

\def\tree{{\rm tree}}

\def\Tr{\, {\rm Tr}}

\def\nn{\nonumber}

\def\eqn#1{eq.~(\ref{#1})}

\def\eqns#1#2{eqs.~(\ref{#1}) and~(\ref{#2})}

\def\NeqFoursYM{{${\cal N}=4$~sYM}}
\def\NeqFour{{{\cal N}=4}}

\def\be{\begin{equation}}
\def\ee{\end{equation}}
\def\bea{\begin{eqnarray}}
\def\eea{\end{eqnarray}}
\def\ba{\begin{eqnarray}}
\def\ea{\end{eqnarray}}

\def\Perm{{\cal P}}

\def\MHVbar{$\overline{\hbox{MHV}}$}

\def\tree{{\rm tree}}
\def\ra{\rightarrow}

\newbox\charbox
\newbox\slabox
\def\s#1{{      
        \setbox\charbox=\hbox{$#1$}
        \setbox\slabox=\hbox{$/$}
        \dimen\charbox=\ht\slabox
        \advance\dimen\charbox by -\dp\slabox
        \advance\dimen\charbox by -\ht\charbox
        \advance\dimen\charbox by \dp\charbox
        \divide\dimen\charbox by 2
        \raise-\dimen\charbox\hbox to \wd\charbox{\hss/\hss}
        \llap{$#1$} }}

\def\subtractfour#1{\ifthenelse{#1=5}{1}{\ifthenelse{#1=6}{2}
{\ifthenelse{#1=7}{3}{\ifthenelse{#1=8}{4}{\ifthenelse{#1=9}{5}
{\ifthenelse{#1=10}{6}{\ifthenelse{#1=11}{7}{\ifthenelse{#1=12}{8}
{\ifthenelse{#1=13}{9}{\ifthenelse{#1=14}{10}{}}}}}}}}}}}

\def\s{\sigma}

\def\MHVbar{$\overline{\hbox{MHV}}$}

\newcommand{\CA}{\mathcal{A}}       
\newcommand{\CG}{\mathcal{G}} 

\newcommand{\CK}{\mathcal{K}}
\def\_{\;\;}
\def\^{\wedge}

\def\Tr{{\rm Tr}}
\def\eqn#1{eq.~(\ref{#1})}

\def\eqns#1#2{eqs.~(\ref{#1}) and~(\ref{#2})}

\def\>{\rangle}
\def\<{\langle}
\def\+{\dagger}
\def\={\ =\ }
\def\tree{{\rm tree}}

\def\and{\qquad\textrm{and}\qquad}

\newcommand{\tvs}{\text{\textvisiblespace}}
\newcommand{\Nf}{{\ensuremath{\mathcal N{=}4}\ }}

\newcommand{\nnl}{\nonumber\\}

\def\FourPointHL#1#2#3#4{
\begin{pspicture}(10.5,8)
\rput[l](1.5,3)
{
\psline{-}(0,0)(7.5,0)\psline{-}(2.5,0)(2.5,2.5)\psline{-}(5,0)(5,2.5)
\rput[Br](-0.4,-0.2){\small{#1}}
\rput[B](2.5,2.9){\small{#2}}
\rput[B](5,2.9){\small{#3}}
\rput[Bl](7.9,-0.2){\small{#4}}
}
\end{pspicture}
}

\def\FivePointHL#1#2#3#4#5{
\begin{pspicture}(13,8)
\rput[l](1.5,3)
{
\psline{-}(0,0)(10,0)\psline{-}(2.5,0)(2.5,2.5)\psline{-}(5,0)(5,2.5)\psline{-}(7.5,0)(7.5,2.5)
\rput[Br](-0.4,-0.2){\small{#1}}
\rput[B](2.5,2.9){\small{#2}}
\rput[B](5,2.9){\small{#3}}
\rput[B](7.5,2.9){\small{#4}}
\rput[Bl](10.4,-0.2){\small{#5}}
}
\end{pspicture}
}

\def\SixPointHL#1#2#3#4#5#6{
\begin{pspicture}(15.5,8)
\rput[l](1.5,3)
{
\psline{-}(0,0)(5,0)\psline{-}(7.5,0)(12.5,0)
\psline[linecolor=red]{-}(5,0)(7.5,0)

\psline{-}(2.5,0)(2.5,2.5)\psline{-}(5,0)(5,2.5)\psline{-}(7.5,0)(7.5,2.5)\psline{-}(10,0)(10,2.5)
\rput[Br](-0.4,-0.2){\small{#1}}
\rput[B](2.5,2.9){\small{#2}}
\rput[B](5,2.9){\small{#3}}
\rput[B](7.5,2.9){\small{#4}} 
\rput[B](10,2.9){\small{#5}}
\rput[Bl](12.9,-0.2){\small{#6}}
}
\end{pspicture}
}

\def\SixPointHLA#1#2#3#4#5#6{
\begin{pspicture}(15.5,8)
\rput[l](1.5,3)
{
\psline{-}(0,0)(5,0)\psline{-}(7.5,0)(12.5,0)
\psline[linecolor=red]{-}(5,0)(7.5,0)
\psline{-}(2.5,0)(2.5,2.5)\psline{-}(10,0)(10,2.5)
\psline{-}(5,0)(5.9,0.9) \psline{-}(6.6,1.6)(7.5,2.5)
\psline{-}(7.5,0)(5,2.5) 
\rput[Br](-0.4,-0.2){\small{#1}}
\rput[B](2.5,2.9){\small{#2}}
\rput[B](5,2.9){\small{#3}}
\rput[B](7.5,2.9){\small{#4}} 
\rput[B](10,2.9){\small{#5}}
\rput[Bl](12.9,-0.2){\small{#6}}
}
\end{pspicture}
}

\def\fork{
\begin{pspicture}(0,0)
\psline{-}(0,0)(0,3)\psline{-}(0,3)(-1.5,4.5)\psline{-}(0,3)(1.5,4.5)
\end{pspicture}
}

\def\forkA{
\begin{pspicture}(0,0)
\psline[linecolor=red]{-}(0,0)(0,3)\psline{-}(0,3)(-1.5,4.5)\psline{-}(0,3)(1.5,4.5)
\end{pspicture}
}

\def\SixPointThreeSym#1#2#3#4#5#6{
\begin{pspicture}(14,8)
\rput[l](7,3.5)
{\rput{0}(0,0){\forkA}\rput{120}(0,0){\fork}\rput{240}(0,0){\fork}
 \rput[Br](-2.9,-5){\small{#1}}
 \rput[Br](-5.2,-1.4){\small{#2}}
 \rput[B](-1.9,5){\small{#3}}
 \rput[B](1.9,5){\small{#4}} 
 \rput[Bl](5.2,-1.4){\small{#5}}
 \rput[Bl](2.9,-5){\small{#6}}
}
\end{pspicture}
}

\def\pssign#1{
\begin{pspicture}(4,8)
\rput[c](2,3.5){#1}
\end{pspicture}
}

\def\HLadderExpand{
\begin{pspicture}(15.5,8)
\rput[l](1.5,2)
{
\psline{-}(0,0)(15,0)
\psline{-}(2.5,0)(2.5,2.5)\psline{-}(5,0)(5,2.5)\psline{-}(12.5,0)(12.5,2.5)\psline{-}(10,0)(10,2.5)
\multido{\ra=6.4+.8}{4}{\psdot[dotsize=3pt,linecolor=blue](\ra,1)}
\rput[Br](-0.4,-0.2){\small{$1$}}
\rput[B](2.5,2.9){\small{$2$}}
\rput[B](5,2.9){\small{$3$}}
\rput[B](9.7,2.9){\small{$m\!\!-\!\!2$}} 
\rput[B](12.5,2.9){\small{$m\!\!-\!\!1$}}
\rput[Bl](15.4,-0.2){\small{$m$}}
}
\end{pspicture}
}

\begin{document}

\preprint{SU-ITP-11/33}
\preprint{NSF-KITP-11-124}

\title{Virtuous Trees at Five and Six Points for Yang-Mills and Gravity}

\author{Johannes Broedel${}^{1,2}$ and John~Joseph~M.~Carrasco${}^{1,2}$}

\affiliation{
${}^1$Stanford Institute for Theoretical Physics and Department of Physics, Stanford University,
Stanford, CA 94305-4060, USA\\
${}^2$Kavli Institute for Theoretical Physics, University of California Santa Barbara, CA 93106, USA}

\begin{abstract}
We present a particularly nice $D$-dimensional graph-based representation of the full color-dressed five-point tree-level gluon amplitude.   It possesses the following virtues: 1) it satisfies the color-kinematic correspondence, and thus trivially generates the associated five-point graviton amplitude, 2) all external state information is encoded in color-ordered partial amplitudes, and 3) one function determines the kinematic contribution of all graphs in the Yang-Mills amplitude, so the associated gravity amplitude is manifestly permutation symmetric.   The third virtue, while shared among all known loop-level correspondence-satisfying representations,  is novel for tree-level representations sharing the first two virtues.   This new $D$-dimensional representation makes contact with the recently found multiloop five-point representations, suggesting all-loop, all-multiplicity ramifications through unitarity.  Additionally we present a slightly less virtuous representation of the six-point MHV and \MHVbar~amplitudes which holds only in four dimensions.
\end{abstract}

\pacs{04.65.+e, 11.15.Bt, 11.30.Pb, 11.55.Bq \hspace{1cm}}

\maketitle


\pagestyle{plain}
\setcounter{page}{1}

\section{Introduction}
In this paper we employ loop-level techniques to establish the existence of tree-level representations satisfying certain  criteria in Yang-Mills (YM) and gravity (GR) theories.  Indeed such criteria expose tree-level calculation to many of the challenges faced in the discovery of particular loop-level representations.   As an eventual goal is to improve at translating between loop-level representations, the existence of such a tree-level proving-ground is convenient.   We will see that there is already practical value for future multi-loop cut calculations arising from this five-point exploration.   Further, study of the form of these representations may help unlock more constructive techniques for satisfying similar criteria at loop-level.

A particularly intriguing discovery is that full color dressed gluon tree-level scattering amplitudes in YM theories encode all the information necessary to write down tree-level graviton scattering amplitudes in related gravity theories.  This was first demonstrated  by Kawai, Lewellen, and Tye (KLT) \cite{Kawai:1985xq} for tree-level open and closed string amplitudes in the eighties.  In the late nineties an all-multiplicity expression was written down by   Bern,  Dixon, Perelstein,  and Rozowsky \cite{BernGaugeGravity} for tree-level field theory.  Recently Bern, Johansson, and one of the current authors (BCJ) discovered that it was possible to extract gravity information from gauge theory representations in a very direct way~\cite{BCJ}.  

To achieve this extraction, one must organize the YM scattering amplitudes into a particularly stringent representation.  First it must be in terms of cubic-vertex graphs, absorbing higher-vertex contact terms into any of the cubic graphs allowed by the color structure.  Secondly, the kinematic factors of the graphs (numerator functions) must be organized so as to share the same algebraic properties as their corresponding color-factors.  Having done so, the gravity amplitude is simply given by a sum over the same cubic graphs, but with a second copy of the Yang-Mills kinematic factor replacing the Yang-Mills color factor, the so called {\it double-copy construction} of gravity amplitudes.   Schematically, if the correspondence is satisfied in the YM representation,
\be
{\rm YM} \propto \!\! \! \sum_{g \in {\rm graphs}} \!\!\frac{n(g) c(g)}{p(g)} \Rightarrow
{\rm GR}\propto \!\!\! \sum_{g \in {\rm graphs}} \!\!\frac{n(g) \tilde{n}(g)}{p(g)},
\ee
where $n(g)$ are the kinematic numerator factors, $c(g)$ are the color factors, $p(g)$ are the propagators of the graphs, and $\tilde{n}(g)$ is simply another copy of the Yang-Mills factor $n(g)$.  The double-copy construction was conjectured and tested to eight points in ref.~\cite{BCJ}, and proven to hold for all multiplicity at tree-level by Bern, Dennen, Huang, and Kiermaier~\cite{LagrangeBCJ}.
 
The story gets even more interesting at the level of quantum (loop-level) corrections: the double-copy construction of gravity amplitudes holds for integrands whenever it is possible to find a YM representation satisfying the color-kinematic correspondence of ref.~\cite{BCJ} as first explicitly demonstrated in \cite{LoopBCJ}.

The four-point loop-level scattering amplitudes in maximally supersymmetric Yang-Mills theory, in representations that allow the double-copy construction of associated gravity amplitudes 
\cite{GSB, BDDPR, LoopBCJ, FourLoopBCJ}, share three virtues:
\begin{enumerate}
\item They satisfy the color-kinematic correspondence of ref.~\cite{BCJ}.  This is sufficient for the double-copy construction.  We will refer to  representations sharing this virtue as  {\it BCJ representations}.  

\item All external state information in the kinematic numerator factors, including any dependence on number of space-time dimensions, is encoded in tree-level color-ordered partial amplitudes.  This virtue carries a number of features.  Their generalization to $D$ dimensions tends to be straightforward. Supersums involving these representations involve only  the partial-amplitude prefactors.  Most importantly, these representations allow for state-agnostic universal expressions.   We will refer to representations sharing this virtue as {\it amplitude-encoded representations}.  

\item Independent of permutations of external leg labels, each graph topology has only one numerator function taking it to kinematic numerators.   This allows for a much smaller number of distinct graph numerator mappings to be specified.   The resulting gravity-expression is then manifestly crossing symmetric.  We will refer to representations sharing this virtue as  {\it symmetric representations}.
\end{enumerate}

Constructing BCJ representations at loop level, while advantageous in terms of minimizing the amount of theory-specific input\footnote{Available most typically through the (generalized) unitarity method~\cite{UnitarityMethod, GeneralizedUnitarity, GeneralizedUnitarity2, BCF}.  See recent reviews for details~\cite{ZviYutinReview, DixonReview, theBestReviewEVAH, BrittoReview}.}, comes with its own challenges.  Namely the process involves solving non-trivial functional relations between graph numerators.  This has been accomplished mainly by introducing sufficiently general ans\"atze~\cite{theBestReviewEVAH,FourLoopBCJ}.   Amplitude-encoding and symmetry, while certainly not necessary at loop level, simplify the finding of BCJ representations by constraining the size of the ans\"atze.  For a particularly impressive example one can consider the representation of four-point four-loop $\NeqFour$ super-Yang-Mills amplitudes in ref.~\cite{FourLoopBCJ}.    This representation allows for the encoding of the entire amplitude in terms of a very small number of  numerator functions.  In particular, all eighty-five symmetric kinematic numerators can be given as functionals of either two planar, or -- even more remarkably -- one non-planar kinematic numerator.   Amplitude encoding allows for a systematic exploration of what ends up being a fairly small ansatz space, as well as recycling the four-loop amplitude in $D$-dimensional box-cuts for all-multiplicity at any higher-loop order. 

At tree-level, on the other hand, finding amplitude-encoded BCJ representations is fairly straightforward~\cite{BCJ}.   Intriguingly, the additional requirement of finding a symmetric representation leads to the type of non-trivial functional relations appearing at loop-level.   This is suggestive because at tree-level all the content of the theory is already available in fairly compact forms -- the color-ordered partial tree-amplitudes.   The discovery of similar challenges as at loop-level means that, in tree-level, we have a potential testing ground for new techniques to move between representations that could be broadly applicable.   

The tree-level BCJ representations  appearing in the literature~\cite{BCJ,  Mafra:2009bz, LagrangeBCJ, KiermaierTalk, BBDSVsolution, Mafra:2011nv,  Mafra:2011nw, Mafra:2011kj, Mafra:2010jq, Mafra:2010gj, Mafra:2010pn, Mafra:2010ir, NSfive, OConnellSelfDual,StringFive} have satisfied, in addition, at most one of the two other virtues, although they may have other very favorable features such as explicit locality in external momenta,  arising naturally from string-theory, compactness, or explicit forms for all multiplicity.  The $D$-dimensional\footnote{The Feynman rules introduced in \cite{LagrangeBCJ} immediately generalize to $D$ dimensions.} tree-level  representations arrived at by the Feynman rules introduced in \cite{LagrangeBCJ} are  symmetric (and local), but at the cost of fairly unwieldy expressions -- embedding external-state information in polarization vectors.     The tree-level representations arrived at by the procedure outlined in \cite{BCJ} are  amplitude-encoded.  Furthermore they make manifest all (generalized) gauge freedom consistent with the correspondence between color and kinematics.    For generic choices of the generalized-gauge, however, these representations are asymmetric;   distinct numerator functions must be defined depending on the permutation of the labeling of the graphs.

In this work we present two new representations of the $D$-dimensional five-point tree -- each sharing all three virtues with the four-point multi-loop representations discussed above.   One of these representations is directly expressible in terms of the universal prefactors appearing in the recently discovered  five-point  multiloop symmetric BCJ representations in \NeqFoursYM~\cite{loop5pt}.  Having identified this relation between tree-level and loop-level, we are able to render the recent five-point multiloop representations amplitude-encoded and thus just as virtuous as the four-point loop-level amplitudes.  Additionally, we present a similar representation for the six-point MHV and \MHVbar~trees that hold only in four dimensions, relying on special four-dimensional properties relating MHV color-ordered scattering amplitudes.

\begin{figure}
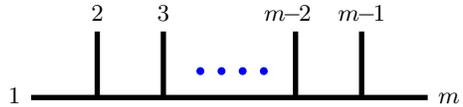

\psset{xunit=10pt,yunit=10pt,linewidth=2pt}
\center{\HLadderExpand}
\caption{Illustration of a generic $m$-point half-ladder diagram.}
\label{halfLadderPicture}
\end{figure}

It is amusing to note that the first and third virtues conspire to engender a particularly satisfying state of grace at tree-level -- the need to specify only one numerator function.   BCJ representations need merely to specify numerator functions for some subset of the half-ladder diagrams~(also termed multi-peripheral diagrams~\cite{HalfLaddersUberAlles}, see \fig{halfLadderPicture}) with various permutations of external leg labels, as all others graph functions are constrained algebraically.  If we can  additionally impose symmetry, we will find that the representation needs only a single function:  all the half-ladders at any given multiplicity will be mapped to kinematic factors with the very same function, but with permuted arguments -- rendering the double-copy constructed gravity amplitudes manifestly permutation symmetric.

The organization of this paper is as follows.  In \sect{Background} we briefly review representations and relations between tree-level partial and fully color-dressed scattering amplitudes, as well as the color-kinematic correspondence of ref.~\cite{BCJ}.   In \sect{Methods} we discuss the general approach to solving the functional constraints of satisfying BCJ, amplitude-encoded, and symmetric representations, and  work out in detail the identification of a virtuous representation of the four-point amplitude.  In \sect{scn:FivePoint} we present and discuss the new representations of the five-point tree, and their relation to higher-loop symmetric-BCJ representations.   In \sect{scn:SixPoint} we  introduce the new representation of the six-point MHV tree. Finally,  in \sect{scn:Conclusion}, we conclude by summarizing the challenges involved with the question of finding similar expressions at higher multiplicity, and discuss the value of exploring various representations at tree-level.

 
\section{Background}
\label{Background}

\subsection{Cubic representation and correspondence between color and kinematics}
The correspondence between color and kinematics relies~\cite{LoopBCJ, theBestReviewEVAH, StructureReview} on the ability to  write any $m$-point $L$-loop, amplitude, with all particles in the adjoint representation, as
\begin{equation}
{(-i)^L \over g_{\text{YM}}^{m-2 +2L }}{\cal A}^L_m = \!\!\!\!\!\!
\sum_{g \in \text{graphs}}{\int{\prod_{i = 1}^L {d^D q_i \over (2 \pi)^D}
\frac{1}{S(g)} \frac {n(g) c(g)}{\prod_{l \in p(g)}{l^2}}}}\,, 
\label{cubicGaugeExpr} 
\end{equation}
where $g_{\text{YM}}$ is the coupling constant.  The sum runs over the set of $m$-point
 $L$-loop graphs with only cubic vertices including permutations of external momenta labels.  
The product in the denominator collects all propagators of each cubic diagram $g$ and 
the integration is performed over all independent loop momenta.   The mapping $c(g)$ takes
the graph $g$ to the color factor obtained by dressing every three-vertex in the graph
with an $\f^{abc} = i \sqrt{2} f^{abc}=\Tr\{[T^{a},T^{b}]T^{c}\}$ structure constant, where 
the color-group generators $T^a$ encode the color
of each external leg $1,2,3 \ldots m$. Accordingly, the mapping $n(g)$
takes the graph $g$ to kinematic factors which can depend on momenta and polarizations -- and can be different for the same topology under permutations of
external leg-momenta. Finally, $S(g)$ denotes the
internal symmetry factors of the individual graphs.
For supersymmetric amplitudes expressed in superspace,
the mapping $n(g)$ will also contain Grassmann parameters. 
The purely cubic form of \eqn{cubicGaugeExpr} can be obtained from other representations by
expressing all contact terms as inverse propagators in
kinematic numerators that cancel propagators.

The correspondence between color and kinematic mappings is satisfied if the kinematic factors obey Jacobi relations in one-to-one correspondence with the color-factors, as well as antisymmetry under the flip of ordering at any odd number of vertices,
\begin{equation}
 n(i)+n(j)+n(k)=0~~ \Leftrightarrow~~ c(i)+c(j)+c(k)=0\,.
\label{Jacobi}
\end{equation}
\begin{equation} 
n(g)\rightarrow -n(\hat{ g} )~ \Leftrightarrow~~c(g) \rightarrow -c(\hat{g} )\,.
\label{vertexFlip}
\end{equation}

 In the early eighties, two papers looking at general gauge theories explored  
 a ``radiation zero'' discovered a few years earlier in an electroweak four-point process~\cite{FourPtBCJ}; the relations they found were
later recognized as the four-point expression of a more general
 color-kinematics correspondence~\cite{BCJ}.  This correspondence (or duality between color and kinematics) holds to all multiplicity at tree-level~\cite{KiermaierTalk,BBDSVsolution}.   
Conjectured to also hold at any loop-level~\cite{LoopBCJ}, the color-kinematic correspondence is  strongly supported in the maximally supersymmetric theories for the four-point amplitudes up through four loops \cite{GSB,BDDPR,LoopBCJ,FourLoopBCJ},  through two loops at five points \cite{loop5pt}, at one loop in ${\cal N}=0\ldots4$~sYM~\cite{BBJ}, and in pure Yang-Mills at two loops~\cite{LoopBCJ}.  Recent reviews of the use of the color-kinematic correspondence in the construction of loop-level amplitudes is given in~\cite{theBestReviewEVAH,
Sondergaard:2011iv}.

\subsection{Color stripped amplitudes and relations}
\label{scn:amplitudeRelations}

In  contrast to \eqn{cubicGaugeExpr}, the full tree-level amplitude can be  alternatively  decomposed,
\begin{widetext}
\begin{equation}
\CA^\tree_m (1,2,3, \ldots, m)=g_{\text{YM}}^{m-2} 
\sum_{\Perm (2,3,\ldots, m)} {\rm Tr}[T^{a_1}T^{a_2} T^{a_3}\cdots T^{a_m}] 
\, A^\tree_m (1,2,3, \ldots, m),
\label{TreeDecomposition}
\end{equation}
\end{widetext}
where $A_m^\tree$ is a tree-level color-ordered $m$-leg partial
amplitude, and the trace is over the group-theory color generators. The sum runs over all noncyclic
permutations of legs, which is equivalent to all permutations keeping
leg $1$ fixed. 

Each color-ordered partial tree-amplitude can in turn be expanded into its subset of the cubic graphs that appear in \eqn{cubicGaugeExpr},
\be
A^\tree_m (1,2,3, \ldots, m)= \sum_{g \in {\rm cyclic}} \frac {n(g)}{\prod_{l \in p(g)}{l^2}}\,,
\ee
where the sum is over all cyclic-relabelings of all topologies that can contribute to the particular color ordering.

Color-ordered tree-level amplitudes satisfy a set of well-known relations. 
The simplest of these are the cyclic and reflection properties,
\bea
A_m^\tree(1,2, \ldots, m) &=& A_m^\tree(2, \ldots, m,1) \,,  \\
\hskip 0.3 cm  A_m^\tree(1,2, \ldots, m) &=& (-1)^m A_m^\tree(m,\ldots,2,1) \nn \,.
\eea
Another relation is  the ``photon''-decoupling identity 
(or subcyclic property)~\cite{ManganoParke, KleissKuijf},
\begin{equation}
\sum_{\sigma \in \rm cyclic} A_m^\tree(1,\sigma(2,3, \ldots, m) ) = 0\,,
\label{Decoupling}
\end{equation}
where the sum runs over all cyclic permutations of legs $2,3,4, \ldots
m$.  This identity follows by replacing $T^{a_1} \rightarrow 1$
in the full amplitude (\ref{TreeDecomposition}), which corresponds to
replacing leg $1$ with a ``photon''.  The amplitude must then vanish
since photons cannot couple directly to particles in the adjoint representation.

Other important relations are the Kleiss-Kuijf
relations~\cite{KleissKuijf}:
\begin{equation}
A_m^\tree(1,\{\alpha\}, m, \{\beta\})=(-1)^{n_\beta}  \!\!\!\!\!\!   \!\!\!\!\!\!    \!\!\!\!\!\! \sum_{\{\sigma\}_i \in {\rm OP}(\{\alpha\}, \{\beta^T\})} 
  \!\!\!\!\!\!   \!\!\!\!\!\!   \!\!\!\! A_m^\tree(1,\{\sigma\} _i,m ) \,,
\label{KleissKuijf}
\end{equation}
where the sum is over the ``ordered permutations'' ${\rm
OP}(\{\alpha\}, \{\beta^T\})$, that is, all permutations of
$\{\alpha\} \bigcup \{\beta^T\}$ that maintain the order of the
individual elements belonging to each set within the joint set.  The
notation $\{\beta^T\}$ represents the set $\{\beta\}$ with the
ordering reversed, and $n_\beta$ is the number of $\beta$ elements.
These relations were conjectured in ref.~\cite{KleissKuijf} and proven in
ref.~\cite{LanceColor}. 
After taking all of the above relations into account, the number of  independent $m$-point amplitudes is $(m-2)!$. 

The ability to construct BCJ representations at tree level  was used~\cite{BCJ} to predict  additional relations between color-ordered partial tree amplitudes, which reduce the number of independent amplitudes to $(m-3)!$. While the general form of the identities is somewhat involved, the structure and the occurrence of kinematic coefficients in the relations can be seen in the following five-point example:
\begin{multline}
s_{24} s_{245} A_5^\tree(1, 2, 4, 5, 3) = -A_5^\tree(1, 2, 3, 4, 5) s_{34} s_{15} \\
  - A_5^\tree(1, 2, 3, 5, 4) s_{14} (s_{245}+s_{35}) \,,
\end{multline}
where $s_{ij\ldots}=(p_i+p_j+\cdots)^2$, and the set of $(5-3)!$ independent five-point tree amplitudes on the right-hand side is obtained by keeping legs $1$ through $3$ fixed.  An all multiplicity expression is given in ref~\cite{BCJ}.
These relations were later derived and proven from string theory using monodromy~\cite{MonodromyNBI1,MonodromyStieberger,MonodromyNBI2}, as well as in a pure field-theoretic approach using on-shell recursion~\cite{BoFengBCJProof1,BoFengBCJProof2}. 

It should be emphasized that all of these relations between partial amplitudes share an important feature:  they hold in arbitrary dimensions.  As such, they can be used to analytically establish $D$-dimensional representations without explicit evaluation in any particular dimension.


\section{Methods}
\label{Methods}
Finding an amplitude-encoded BCJ satisfying representation at $m$-point tree-level is straightforward.  We start  by identifying the cubic tree graphs with $m$ external legs,  independent under vertex-flip antisymmetry, and write down the linear system of equations generated by the Jacobi relations between kinematic numerator factors.     We can reduce this linear system by simple elimination of kinematic  factors, solving each  in terms of simple linear combinations of others, until no more elimination is possible, and we are left with a solution for every kinematic factor as a linear functional of the graphs independent under these relations.  These independent  graphs are termed  ``master graphs'', as they effectively encode the full amplitude.    It is important to realize that these  master graphs are often related by  graph isomorphisms, their `independence'' is only under the Jacobi relations.  As such, the same topology may appear several times with different labelings in the master graphs. 

We can take any set of independent partial amplitudes, decompose them into their cubic-graph representation, and express their kinematic factors in terms of the master kinematic factors.    As there are $(m-3)!$ independent partial amplitudes for $m$-point interaction, this allows us to solve for $(m-3)!$ of the master kinematic factors in terms of the independent partial amplitudes, propagators, and the remaining unconstrained kinematic factors associated with the other master graphs.  

At this point we have a complete BCJ, amplitude-encoded representation:   all external dependence of the scattering amplitude are encoded in the $(m-3)!$ color-ordered partial amplitudes, and the representations satisfy the color-kinematic Jacobi relations by construction.    None of the unconstrained factors can affect the actual value of the scattering amplitudes if the constrained $(m-3)!$ numerator kinematics have been defined as above, so they are described as parameterizing a generalized gauge freedom~\cite{BCJ}.     These dynamic\footnote{Functions of kinematics.}  parameters can be set to any value.  They could be set to vanish, or chosen to be functions that maximize the total number of graphs whose numerators vanish, c.f.  the  representations of ref.~\cite{KiermaierTalk,BBDSVsolution}.

Recall that many of these graphs may share the same topology -- in fact, the master graphs can all be chosen to be various labelings of the half-ladder\footnote{This was demonstrated by Del Duca, Dixon, and Maltoni for tree-level color-factors \cite{HalfLaddersUberAlles}.  As the argument was based purely based upon Jacobi identities it holds for kinematic numerators in a BCJ representation.}-- as we will do for the rest of this discussion.  For symmetric representations, each of these master numerators $n(g)$ will be given by the same function $n_{\rm hl}$ taking as its argument the various permutations of external labels.   All decompositions of color-ordered partial amplitudes in terms of their cubic graphs simply represent functional constraints  that the $n_{\rm hl}$ must satisfy.   The Jacobi relations and symmetry relations reflect the sole remaining functional constraints.  Finding an ansatz general enough to satisfy these functional constraints, and yet remain computationally tractable, poses the primary obstacle to finding symmetric BCJ representations at tree-level.

\begin{figure}[t]
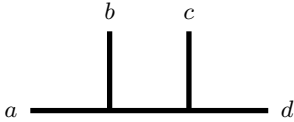

\psset{xunit=12pt,yunit=12pt,linewidth=2pt}
\center{\FourPointHL{$a$}{$b$}{$c$}{$d$}}
\caption{The four point cubic diagram.   It appears with three distinct labelings of external legs $(a,b,c,d)$, corresponding to the ``$s$''-channel diagram: $(1,2,3,4)$, the ``$t$''-channel diagram: $(2,3,4,1)$, and  the ``$u$''-channel diagram: $(3,1,4,2)$.    }
\label{fourPointFig}
\end{figure}

\subsection{Four-point example}
Let us warm up by seeing how this all comes together at four points.   There is only one cubic topology at four points, shown with arbitrary labels $(a,b,c,d)$ in \fig{fourPointFig}.  We first consider all 24 ways of giving $(a,b,c,d)$ different values of $(k_1, k_2, k_3, k_4)$.  If we recognize that there is nothing to distinguish the labeled graph $(a,b,c,d)$ from $(c,d,a,b)$, then the number of labelings drops to 12.  Imposing antisymmetry, $ n(b,a,c,d)=-n(a,b,c,d) $,  takes us to only 6 labelings, and considering the same flip on the other vertex reduces the number of distinct labelings to three.  These three four-point graphs are traditionally named for the Mandelstam variable carried as their propagators ($s=(k_1+k_2)^2$, $t=(k_2+k_3)^2$, and $u=(k_3+k_1)^2$ with $s+t+u=0$).  We will use the shorthand notation $n_s$, $n_t$, and $n_u$, for their three different numerator functions.   Here, and as in the rest of the paper, external state indices on external particles are suppressed -- they are to be taken as to follow the momentum labels in any argument to numerators or color-ordered partial amplitudes.   The cubic representation of the full four-point amplitude is then
\be
\label{full4pt}
{\cal A}^\tree_4 = g_{\text{YM}}^2 \left( \frac{n_s c_s}{s} + \frac{ n_t c_t}{t} + \frac{n_u c_u}{u} \right) \,,
\ee
where $g_{\text{YM}}$ is the gauge coupling constant, $c_i$ are the associated color factors with each tree-graph.  There is but one kinematic Jacobi relation between these graphs, and following the signs of the color factors associated with this edge ordering, it is $n_u=n_s-n_t$, in correspondence with the color Jacobi relation $c_u=c_s-c_t$.

We can write down any $(4-3)!=1$ independent color-ordered partial amplitude, which without loss of generality we choose to be 
\be
A^\tree_4(1,2,3,4)= n_s/s + n_t/t,
\label{stTree}
\ee
and solve it for one of the numerators.  Now we can express $n_t$ as a function of of $n_s, s,t,$ and $ A^\tree_4(1,2,3,4)$,
\be
\label{fourBAsoln}
n_t= t \,\Bigl( A^\tree_4(1,2,3,4)-\frac{n_s}{s} \Bigr),
\ee
and together with
\be
n_u = n_s -n_t =- \frac{u}{s}\,n_s - t \, A^\tree_4(1,2,3,4)\,,
\ee
we have a full BCJ, amplitude-encoded solution.  With a little algebra one can see that $n_s$ drops out of all physical quantities, and so parameterizes the generalized gauge freedom consistent with a BCJ representation. 
 
As mentioned, all external state information is encoded in the color-ordered scattering amplitude in the solution to $n_t$, and all residual generalized-gauge freedom is encoded in $n_s$.  For example we are free to choose $n_s=- \bigl(\frac{s\,t}{u}\bigr) A^\tree_4(1,2,3,4) $ such that $n_u$ vanishes.  
Similarly, one could have chosen $n_s$ such that $n_t$ explicitly vanishes, or simply set $n_s=0$.  Unlike, for example, the non-dynamic single-parameter gauge freedom we will find in our symmetric five-point representation, this asymmetric freedom is much more flexible: $n_s$ can be any function whatsoever.  The natural question to the point of this paper is whether one can find a symmetric representation: i.e. the same function $n(a,b,c,d)$ such that   $n(1,2,3,4)$ returns an appropriate $n_s$,    $n(2, 3,4,1)$   returns an appropriate $n_t$, and   $n(3,1,4,2)$ returns an appropriate $n_u$.

It will turn out to indeed be possible. In order to find the correct form of $n(a,b,c,d)$ we need to solve the following functional constraints:
\begin{eqnarray}
\label{jacobiConstraint}
 n(a,b,c,d) &=& n(c,a,d,b) + n(b,c,d,a) \\
 \label{decompositionConstraint}
A^\tree_4(a,b,c,d) &=& \frac{n(a,b,c,d)}{s_{ab}} +\frac{n(b,c,d,a)}{s_{bc}}\\
n(a,b,c,d)&=& ~n(c,d,a,b)\\
\label{symmConstraints}&=&- n(a,b,d,c)\nnl
&=& - n(b,a,c,d)\nonumber\,,
\end{eqnarray}
where the first relation is the Jacobi identity $n_s=n_u+n_t$, the second is the decomposition of the color-ordered partial tree amplitude, and the last three impose the required graph autmorphisms.

We introduce an ansatz for the form of $n(a,b,c,d)$.  In order to satisfy amplitude encoding we will express our ansatz in terms of color-ordered partial amplitudes.  Given that $(4-3)!=1$, we could choose to use an ansatz that only depends on a single color-ordered scattering amplitude. As the other partial amplitudes will be related to that  by ratios of momentum invariants, however,  our intermediary stages would be a little more complicated.  Rather we choose an ansatz involving two  amplitudes $A^\tree_4(a,b,c,d)$ and $A^\tree_4(a,c,b,d)$ which span the color-ordered amplitude space without the need to put  any $s_{ab}$ in  denominators.  

Recognizing that 
$$ 
s_{ac} A^\tree_4(a,c,b,d) =s_{ab} A^\tree_4(a, b, c, d)\,
$$ and  that the sum of the mandlestam invariants vanish, our simplest  ansatz satisfying the dimensionality requirements is
\begin{multline}
n(a,b,c,d)  = \alpha ~ s_{ab} A^\tree_4(a,b,c,d)  \\
+ \beta  ~  s_{ad} A^\tree_4(a,b,c,d)  
+ \gamma  ~s_{ad} A^\tree_4(a,c,b,d),
\end{multline}
where $\alpha, \beta, \gamma$ are scalars that should  satisfy the constraints of \eqn{jacobiConstraint}, \eqn{decompositionConstraint}, and \eqn{symmConstraints}.  Carrying out the calculation one finds the parameters to be completely fixed to $\alpha=1/3,\,\beta=0,\,\gamma=-1/3$ giving a symmetric, BCJ, amplitude-encoded representation.

Replacing $A^\tree_4(a,c,b,d) = A^\tree_4(a,b,c,d) s_{ab}/s_{ac}$ in our solution, we find something quite striking:
\bea
\label{evocativeFour}
n(a,b,c,d) & =&  \frac{ A^\tree_4(a,b,c,d)}{s_{ac}} \frac{1}{3} s_{ab}  (s_{ac} - s_{bc})\\
&=& \frac{[ s_{ab} s_{bc} A^\tree_4(a,b,c,d) ] }{s_{ab} s_{bc} s_{ac}} \frac{1}{3}s_{ab}  (s_{ac} - s_{bc})\nn\, .
\eea
The most notable thing about this form is that the numerator of the first fraction (in the square brackets) is the  universal prefactor $\CK_4$~\cite{Neq44np} of the four-point multiloop $\NeqFour$~sYM amplitudes  used to encode all external state information.  It is invariant under permutations between leg labels, as is the denominator of that fraction.   This means that all the antisymmetry properties of the numerator must be satisfied by the function of momentum invariants.  Indeed $s_{ab} ( s_{ac} - s_{bc} )$ is the simplest function of momentum invariants that satisfies the symmetry properties of \eqn{symmConstraints}, and could, in principle, be guessed ahead of time.  The denominator $s_{ab} s_{bc} s_{ac}$  is proportional to the Gram determinant ${\cal G}_4$ relevant to four-point interaction, i.e.  ${\cal G}_m \equiv \det(  k_i \cdot k_j )$, where for $m=4$, $i$ and $j$ run from 1 to 3.

The form in \eqn{evocativeFour} is  evocative as a starting point for the types of expressions that might generalize to symmetric higher points.  Recalling that $\CG_m$ goes as $s^{(m-1)}$, and there are $(m-3)$ propagators in any cubic tree diagram,  we can arrive at the idea that the $D$-dimensional $m$-point half-ladder may be schematically of the form 
\bea
\label{genericAnsatz}
n_{m,{\rm hl}}\propto \sum \alpha \bigl[ \frac{s^{m-1}}{\CG_m}  A^\tree_m s^{m-3} \bigr],
\eea
where the sum will be over all $(m-2)!$ Kleiss-Kuijf independent color-ordered partial tree-amplitudes as well as all independent $s^{2m-4}$-order products of momentum invariants $s_{ij}$, and the $\alpha$ represent parameters to be constrained by the relevant symmetries, Jacobi-identities, and amplitude equations.

One can see that using \eqn{genericAnsatz} as an ansatz for actually solving the functional relations is prohibitive. As it grows quickly in the number of external particles, is not particularly practical even at five points. Fortunately, we can begin our exploration at five points with the simpler type of polynomial ansatz we started with at four points, and see where we need to enlarge to include rational terms so as to relate to the recent higher-loop results of ref.~\cite{loop5pt}.


\section{Five-point tree}
\label{scn:FivePoint}
\subsection{First representation}

\begin{figure}
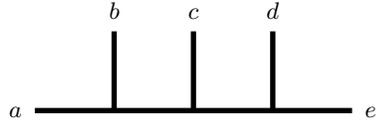

\psset{xunit=12pt,yunit=12pt,linewidth=2pt}
\center{\FivePointHL{$a$}{$b$}{$c$}{$d$}{$e$}}
\caption{The five point half-ladder diagram.  All contributions in cubic-graph representations of five-loops involve this topology or are related by antisymmetry around vertices. }
\label{halfLadder5}
\end{figure}

Considering now five points, there is again only one graph topology, the half-ladder depicted in \fig{halfLadder5}. For this topology, there are fifteen distinct labelings under vertex antisymmetry. 
Similar to our approach with four points we may start with an ansatz comprised of color-ordered partial scattering amplitudes with two powers of momentum invariants $s_{ij}$ to cancel the two propagators in the cubic graphs. Since there are only five independent $s_{ij}$  for massless five-point amplitudes and only six linearly independent Kleiss-Kuijf independent amplitudes, this ansatz is quite manageable.   We solve using the constraints  of the symmetry and Jacobi relations.  It is worth noting that only a single color-ordered tree-level cubic-graph  decomposition is required as a functional constraint, e.g.,
\begin{multline}
A^\tree_5(1,2,3,4,5) =
\frac{1}{s_{1 2} s_{4 5}} n_5(1,2,3,4,5)\\+
\frac{1}{s_{2 3} s_{1 5}} n_5(2,3,4,5,1)+
\frac{1}{s_{3 4 } s_{1 2}} n_5(3,4,5,1,2)\\
+\frac{1}{s_{4 5 } s_{2 3}} n_5(4,5,1,2,3)+
\frac{1}{s_{1 5  } s_{3 4}} n_5(5,1,2,3,4)\, .
\label{fivePointTreeConstraint}
\end{multline}
The numerator function must obey the following graph symmetry relations:
\bea
\label{fivePointSymm}
n_5(a,b,c,d,e)&=& - n_5(b,a,c,d,e)\\
&=& -n_5(a,b,c,e,d)  \nn \\
&=&-n_5(e,d,c,b,a) \nn \, ,
\ea
and the following two Jacobi identities
\bea
\label{fivePointJacobi}
n_5(a,b,c,d,e)\!\! &=&\!\! n_5(d,e,a,b,c) + n_5(d,e,b,c,a)\\
n_5(a,b,c,d,e)\! \!&=&\!\! n_5(a,b,e,d,c) + n_5(e,c,d,a,b)\nn \,.
\ea

These constraints are sufficient to ensure a correct reproduction of the full amplitude.   Solving these relations we indeed find a $D$-dimensional solution with no additional freedom left in the ans\"atze.  The form is flexible in the sense that various color-ordered amplitudes are related to each other under the relations discussed in \sect{scn:amplitudeRelations} -- but the result is unique.   One nice expression of this numerator function is as follows, 
\begin{multline}
\label{rep51}
n_{5,1}(a,b,c,d,e)=
\frac{1}{30} \Bigg(\\
\Big[s_{ab}s_{de}(A_{a b c d e}-A_{a b c e d}
-A_{b a c d e}+A_{b a c e d})\Big]\\
\!\!\!\!\!\!\!\!\!\!\!\!\!\!\!\!\!\!\!\!+ \Big[s_{ab}(s_{cd}-s_{ce})(A_{a d c e b}+A_{a e c d b})\\
+
s_{de}(s_{ac}-s_{bc})(A_{e a c b d}-A_{d a c b e})\Big]\\
\!\!\!\!\!\!\!\!\!\!\!\!\!\!\!\!\!\!\!\!+\Big[(s_{ab}s_{cd}-s_{ab}s_{ce})A_{a d c e b}+(s_{ab}s_{cd}-s_{ab}s_{ce})A_{a e c d b}\\
+(-s_{ae}s_{bc}-s_{be}s_{cd})A_{a d c b e}+(s_{ad}s_{bc}+s_{bd}s_{ce})A_{a e c b d}\\
~~~~~+(s_{ac}s_{bd}+s_{ad}s_{ce})A_{d a c e b}+ (-s_{ac}s_{be}-s_{ae}s_{cd})A_{e a c d b}\Big]
\Bigg)
\end{multline}
where we introduce the notation
$$
A_{abcde}\equiv A^\tree_5(a,b,c,d,e)\,.
$$ 
The solution \eqn{rep51} is not the most compact available expression, but it makes the automorphism symmetries of \eqn{fivePointSymm} quite manifest as we will discuss.
 
Indeed, a first guess as to the answer  might be to make use of the reflection properties of the partial amplitudes and combine them in such a way so as to incorporate the other constraints of \eqn{fivePointSymm}.  From there, we  need only multiply by a pair of momentum products, $s_{ab} s_{de}$, invariant under the antisymmetries, to arrive at
\begin{equation}
s_{ab}s_{de}\left(A_{a b c d e}-A_{a b c e d}-A_{b a c d e}+A_{b a c e d}\right)\,.
\end{equation}
This simple expression, the first block of \eqn{rep51}, while satisfying all the symmetry constraints, fails to satisfy  \eqn{fivePointJacobi}, the functional Jacobi relations. A little more thought about various ways of representing the antisymmetry constraints, may lead  to each of the other two blocks appearing in \eqn{rep51}.  Each of these independently satisfies the antisymmetry conditions.  Combining all three with the correct prefactor to solve 
\eqn{fivePointTreeConstraint}, also solves the Jacobi relations, and so \eqn{rep51} is the solution that satisfies all of our desired virtues.

Using the $D$-dimensional relations allowing one to express every five-point color-ordered  amplitude in terms of $A^\tree_5(1,2,3,4,5)$ and $A^\tree_5(1,2,3,5,4)$, and conservation of momentum to relate the $s_{ij}$, it is straightforward to verify that this respects all the constraints, and so generates a $D$-dimensional  representation,
\be
\label{fiveYM}
{\cal A}^{(0)}_5 = g_{\text{YM}}^3 \!\!\!\!\! \sum_{\{q_1,...,q_5\} \in S_5} \!  \frac{1}{8} \frac{ c(q) n(q)}{p(q)},
\ee
where $g_{\text{YM}}$ is the gauge coupling constant, $c(q)$ comes from dressing the half-ladder with appropriate color factors for the labels $q$,  $p(q)$ gives the product of the propagators associated with that labeling, and we simply sum over all permutations of external leg labels, dividing by an overall symmetry factor, in this case $8$.  In this case we take $n$ to be $n_{5,1}$.   Given the satisfaction of the color-kinematic correspondence (the first virtue) we can trivially write down the known gravity amplitude in a manifestly crossing symmetric representation:
\be
\label{fiveGR}
{\cal M}^{(0)}_5 = i \Bigl(\frac{\kappa}{2}\Bigr)^{3} \!\!\!\!\! \sum_{\{q_1,...,q_5\} \in S_5} \!  \frac{1}{8} \frac{ n(q) n(q)}{p(q)}.
\ee

These are fine amplitude representations that satisfy all of our requirements.  Familiar with the beautiful structure relating four-point tree and multi-loop corrections in ${\cal N}=4$~sYM, there is one additional property one may desire from a five-point tree-level representation: that it be manifestly constructed with the same building blocks that appear in higher-loop same-multiplicity amplitudes.  In other words, it would be nice to relate to the multiloop structure that appears in five-point ${\cal N}=4$ sYM~\cite{loop5pt}.  We will take a brief detour to review this newly discovered five-point multiloop structure, and go on to find a second, entirely distinct, symmetric numerator function for the tree-level five-point half-ladder.

\subsection{Multiloop structure and second representation}

It was recently shown \cite{loop5pt} that the five-point one- and two-loop amplitudes in \Nf sYM possess a very compact structure if written in a symmetric BCJ representation, where the external legs are in four-dimensions, but all loop-momenta are allowed to run in $D$ dimensions.  The one-loop master-numerator (labeled with consecutively increasing legs) has the following cyclically symmetric form 
\be
n^{(1)}_{\rm pentagon} = \beta_{12345}\equiv \delta^{(8)}(Q)\frac{  \spb{1}.{2} \spb{2}.{3} \spb{3}.{4} \spb{4}.{5} \spb{5}.{1} }{4 \, \varepsilon(1,2,3,4)}\,,
\ee
where the external states are packaged in the usual Grassman delta function $\delta^{(8)}(Q)$.  The denominator is the Levi-Civita invariant,  $\varepsilon(1,2,3,4)\equiv \varepsilon_{\mu\nu\rho\sigma}k_1^\mu k_2^\nu k_3^\rho k_4^\sigma$.   For compactness the authors of ref.~\cite{loop5pt} also introduce the following functions,
\be
 \gamma(ijklm) \equiv \beta(ijklm)-\beta(jiklm),
\ee
which are antisymmetric in $ij$, but symmetric  in $klm$, such that the last three arguments can be suppressed $\gamma_{ij} = \gamma(ijklm)$. These $\beta$ functions encode all five-point external state information through two loops, and are conjectured to extend to all loops. The remaining components of the numerator factors are constrained to be monomials in Lorentz products of the appropriate engineering-dimension weight.  In fact, in \cite{loop5pt} it was suggested that the tree-level numerators could be expressed in the following functional form
\be
\label{CJconj}
n^{(0)} \sim \sum  \beta(\tvs)/s_{\tvs},
\ee
i.e. some linear combination of the $\beta$ functions divided by a momentum invariant $s_{ab}$.  This is not the representation found in \eqn{rep51}.  In order to make this clearer, we can rewrite $\beta$ as follows
\begin{multline}
\beta(1,2,3,4,5) = -i \,{A^\tree_5(1,2,3,4,5)} \\
\times {s_{12}s_{23} s_{34} s_{45} s_{51}}  \frac{  \varepsilon(1,2,3,4)}{4\,{\CG}_5}\,,
\end{multline}
where we have used the Parke-Taylor representation of the superamplitude to absorb the spinor-products and Grassmann delta function. Furthermore we multiplied the numerator and the denominator by $\varepsilon$ in order to introduce $\CG_5=-\varepsilon^2$, the Gram determinant relevant to five-point massless interaction. This expression diverges when the momenta are restricted to a three-dimensional subspace.
Thus there is no way to describe $\beta$ in terms of linear combinations of $n_{5,1}$,
\be
\forall~ \alpha: \hskip1cm \beta(1,2,3,4,5) \ne \!\!\!\!\!\sum_{ \{q_1,...,q_5\} \in S_5}\!\!\!\!\! \alpha_q ~ n_{5,1}(q)) \,.
\ee
Does this mean that the tree-level numerator cannot be described by something of the form ~\eqn{CJconj}?  Not at all.  This simply means that the representations given in \eqn{rep51} and suggested in \eqn{CJconj} are distinct.    

Naturally we now consider an ansatz of the form \eqn{CJconj}.  Using again, symmetry, the Jacobi relations, and the single graph decomposition \eqn{fivePointTreeConstraint}, we find that that such an ansatz can, independent of $n_{5,1}$, also satisfy all constraints.   Given the form of $\beta$ these verifications must be performed explicitly in four dimensions, and are most easily done numerically.    The form of this numerator function is completely constrained (other than trivial relations between $\beta$ functions with differently permuted arguments),
\begin{multline}
\label{rep52}
n_{5,2}(a,b,c,d,e)=
\frac{1}{10} \Bigg( \Big[ \left(\frac{1}{s_{c d}}-\frac{1}{s_{c e}}\right) \gamma_{a b} \Big]\\
   + \Big[ \left(\frac{1}{s_{a c}}-\frac{1}{s_{b c}}\right) \gamma_{ed} \Big]\\
- \Big[ \frac{\beta_{e d c b a  }}{s_{a e}}+\frac{\beta_{d e c a b  }}{s_{b d}}
-
\frac{\beta_{ e d c a b  }}{s_{b e}}-
\frac{\beta_{d e c  b  a  }}{s_{a d}} \Big] \Bigg)\,.
\end{multline}
Thus we have found an entirely different representation of the five-point tree-level amplitude that is at least as virtuous as the first.   

Now, as both solutions accurately describe the five-point color-ordered trees (one can check that the divergence in $\beta$ for dimensionally restricted subspace cancels between the numerators), one can solve for $\beta(1,2,3,4,5)$ entirely in dimension-agnostic terms:
\begin{multline}
\label{goodBeta}
\beta^{D}(1,2,3,4,5) \equiv \frac{s_{12}s_{23} s_{34} s_{45} s_{51}}{16\, {\cal G}_5} 
\Bigg[\\
\left(s_{1 5} s_{3 4}+s_{1 4} s_{3 5}-s_{1 3} s_{4 5}\right)
   A^{\text{tree}}\left(  1, 2, 3, 4, 5\right) \\
 +  2 s_{1 4} s_{3 5}
   A^{\text{tree}}\left(  1, 2, 3, 5, 4\right)\Bigg] \, .
\end{multline}
This is a particularly appealing form, as state-sums involving 5-point amplitudes of any loop-level have significant   generalized unitarity-cut ramifications, and this reduces the question to  state-sums involving five-point trees, known in the maximally supersymmetric case from the three-particle cut of the four-point two-loop amplitude given in ref.~\cite{BDDPR}.

Taking $\beta \to \beta^{D}$  in \eqn{rep52}, we have checked that $n_{5,2}$ satisfies all the $D$-dimensional constraints, i.e. that under the $D$-dimensional relations algebraically relating all color-ordered tree-amplitudes to a basis of $(5-3)!=2$ color-ordered amplitudes, and conservation of momentum, it correctly reproduces all color-ordered partial tree-amplitudes decomposed into cubic graphs.  As such \eqn{rep52} generates another $D$-dimensional BCJ, amplitude-encoded, symmetric five-point representation of Yang-Mills and gravity, when used in \eqn{fiveYM} and \eqn{fiveGR} respectively.   As $n_{5,1}$ and $n_{5,2}$ are distinct, one can parameterize the gauge freedom with a single complex parameter,
\be
n_{5} \equiv \alpha \, n_{5,1} + (1-\alpha) \,n_{5,2}\,.
\ee
This  is consistent with the known single-parameter gauge-freedom of the symmetric, BCJ (non-amplitude-encoded) representation found in ref.~\cite{LagrangeBCJ}.

It is worth highlighting that the form of $n_{5,2}$ in \eqn{rep52} is a manifestation of the type of expression sketched in \eqn{genericAnsatz}.    Attempting to directly fit \eqn{genericAnsatz} to the data even at five points would have been somewhat laborious.   The functional building block, $\beta$, was instead arrived at by looking at the maximal cut of the five-point one-loop amplitude in $\NeqFour$~sYM.  

\section{Six-point MHV tree in four dimensions} 

\label{scn:SixPoint}

\begin{figure*}
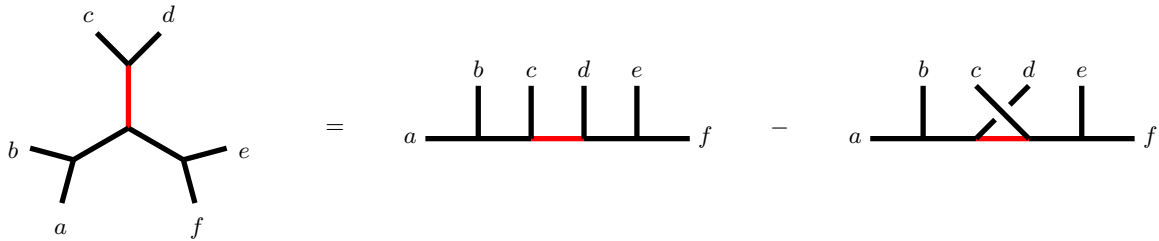

\psset{xunit=8pt,yunit=8pt,linewidth=2pt}
\center{\SixPointThreeSym{$a$}{$b$}{$c$}{$d$}{$e$}{$f$}\pssign{$=
$}\SixPointHL{$a$}{$b$}{$c$}{$d$}{$e$}{$f$}\pssign{$-$}\SixPointHLA{$a
$}{$b$}{$c$}{$d$}{$e$}{$f$}}
\vskip.2cm
\caption{Illustration of the kinematic Jacobi relation associated with the indicated edge, as given in \eqn{sixJacRule}, which expresses the numerator of the trimerous topology on the left in terms of the difference between the two half-ladders on the right.}
\label{sixPointRelation}
\end{figure*}

In this section we introduce  a  four-dimensional symmetric, amplitude-encoded, BCJ representation of the six-point tree diagram.   
Contrary to the lower-point trees, there are  two distinct topologies contributing to the six-point tree amplitudes; in addition to the half-ladder we now have trimerous graphs (as depicted in \fig{sixPointRelation}).  As mentioned earlier, we can express the second topology in terms of the first via
\begin{multline}
\label{sixJacRule}
n_{6,\rm tri}(a,b,c,d,e,f)= n_{6,{\rm hl} }(a,b,c,d,e,f) \\
-n_{6,{\rm hl}}(a,b,d,c,e,f) \,,
\end{multline}
so we need only concern ourselves with finding $n_{6,{\rm hl}}$.   Beyond five points,  even ans\"atze merely polynomial in momenta invariants grow rapidly in size.  Furthermore, it is not so clear whether such a form is sufficient to capture the behavior that generalizes to higher loops. As seen at five points, functions which generalize to loops can involve rational functions of momentum invariants and not just polynomials.  

For this work, we have not carried out the exploration of $n_{6,{\rm hl}}$ systematically. Rather, we guessed at a form using a small number of color-ordered partial amplitudes, sufficient to allow a na\"ive symmetry encoding, multiplied by the denominators that appear in their cubic-graph expansion.   We fit this ad-hoc ansatz to the Jacobi relations, the symmetry-constraints, and the  cubic-graph decomposition of the partial-amplitude.  Such an intuitive approach  yielded a valid expression but with some -- perhaps not so surprising -- limitations.  The following compact expression for $n^6_{\rm hl}$ only holds in four dimensions, and only holds for MHV and \MHVbar amplitudes as it relies on special four-dimensional identities,
\begin{multline}
\label{sixFun}
 n_{6,{\rm hl}}( a, b, c, d, e, f)=
\frac{s_{ a  b}}{15}
~\Big(- s_{ d  c} s_{ e  f} A_{ a   b   d   c   e   f}\\
+  s_{ d  c} s_{ f  e} A_{ a   b   d   c   f   e }
-  s_{ d  f} s_{ e  c} A_{ a   b   e   c   d   f }
 -  s_{ c  f} s_{ e  d} A_{ a   b   e   d   c   f }\\
-  s_{ c  d} s_{ e  f} A_{ a   b   e   f   c   d } 
+  s_{ d  e} s_{ f  c} A_{ a   b   f   c   d   e }
+  s_{ c  e} s_{ f  d} A_{ a   b   f   d   c   e } \\
+  s_{ c  d} s_{ f  e} A_{ a   b   f   e   c  d }\Big)\, ,
\end{multline}
where $A_{abcdef}\equiv A^\tree_6(a,b,c,d,e,f)$.

Under those limitations it does generate the appropriate symmetric, BCJ, amplitude-encoded representations of Yang-Mills and gravity theories respectively,
\be
\label{sixYM}
{\cal A}^{(0)}_6 = g^4 \sum_{q \in S_6} \Bigl( \frac{1}{8}  \frac{ c_{\rm hl}(q)~ n_{6,{\rm hl}}(q)}{p_{\rm hl}(q)}+
 \frac{1}{48}  \frac{ c_{\rm tri}(q) ~ n_{\rm tri}(q)}{p_{\rm tri}(q)} \Bigr)\,
 \ee  
\be
\label{sixGR}
{\cal M}^{(0)}_6 = i \Bigl(\frac{\kappa}{2}\Bigr)^{4}  \sum_{q \in S_6} \Bigl( \frac{1}{8}  \frac{ (n_{6,{\rm hl}}(q))^2}{p_{\rm hl}(q)}+
 \frac{1}{48}  \frac{  ( n_{\rm tri}(q))^2}{p_{\rm tri}(q)} \Bigr)\,.
 \ee
 The factors of 8 and 48 are the symmetry factors of the half-ladder and trimerous graphs.   One sees that, as before, the gravity amplitude is manifestly permutation symmetric.

It should be stressed that the limitations of this representation does not reflect any tension between BCJ representations and non-MHV amplitudes.   Indeed, the  all-multiplicity  amplitude-encoded BCJ representations in the literature hold in any dimensions, independent of external states.  The struggle is to find an ansatz general enough to allow for the solution of the functional constraints, and at the same time being computationally tractable.  It is easy to believe that a form of the type \eqn{genericAnsatz} may work in $D$ dimensions, independent of external states, but the most direct path to reveal it seems to await a better understanding of the structures involved.



\section{Conclusions}
\label{scn:Conclusion}

 After we worked out the four-point  BCJ, amplitude-encoded, symmetric numerator in explicit detail, we presented two independent five-point  $D$-dimensional representations, one of which is related to the structure recently uncovered at multi-loop five-point in the maximally supersymmetric theory.  Exploring the consequences of these two representations,  we rendered, {\it en passant}, the five-point multiloop amplitudes as virtuous as the four-point multi-loop amplitudes by finding an amplitude-encoded form of the $\beta$ function, \eqn{goodBeta}.  In effect, this relates, in the maximally supersymmetric theory,  the state-sum of all three-particle cuts involving two five-point sub-amplitudes,  to the known three particle cut of the four-point two-loop amplitude.  We also presented a slightly less virtuous six-point representation.

An obvious goal is to identify a constructive principle for virtuous representations.  The underlying kinematic algebra responsible for the color-kinematic correspondence, however, is unknown beyond certain sectors in four dimensions~\cite{OConnellSelfDual}.   The existence of such an algebra is suggested in general, not only by the kinematic Jacobi relations, but additionally by a trace basis identified by Bern and Dennen in ref.~\cite{TraceDual}.  They present an alternative amplitude representation based on swapping the role of color and kinematics in the traditional color-trace decomposition of \eqn{TreeDecomposition}.   The partial-amplitudes in their representation involve the color-factors as numerators, and they introduce kinematic ``traces'' $\tau(q_1\ldots q_m)$ in place of the trace over color-generators.   It is perhaps worth noting that the numerator functions presented here, for five and six points, each lead to a symmetric  $\tau$, i.e. a single function which takes any labeling to the appropriate kinematic contribution.

The appeal of  BCJ representations at loop-level resides in the ability to propagate a minimal amount of information from the theory into the full amplitude~\cite{LoopBCJ, loop5pt,  FourLoopBCJ}, as well as the ability to trivially generate loop-level gravity amplitudes.  At tree-level, where representations are already known for both Yang-Mills and gravity theories, the appeal is more subtle.  Namely it resides in the relations satisfied by the numerators.  Expansion  of  tree level numerators in terms of color-ordered partial amplitudes might arguably seem to import to Yang-Mills the aesthetics of KLT gravity representations rather than vice-versa.  Here, however, we show that these (less than compact) expressions  do manage to serve an aesthetic ideal, they make kinematic symmetries manifest in concordance with the color-factor symmetries.   The  introduction of combinations of partial trees which satisfy these symmetry properties (like the separate symmetry-respecting blocks in \eqns{rep51}{rep52}),  lead to quite natural $D$-dimensional expressions.   It is to be hoped that by studying the different possible representations that exist at tree-level,  one may discover  novel ways of moving between representations that generalize to loop-level.

We see that the tree-level requirement of symmetric BCJ representations involves the solution of non-trivial functional relations analogous to the operations necessary for finding loop-level BCJ representations.   In  
\eqn{genericAnsatz} we sketched an ansatz for numerators based upon the types of tree-level structures that generalize to higher loops for the known multiloop BCJ representations in $\NeqFour$~sYM at four and five points.   It is clear that there are ways of packaging color-orderd partial amplitudes that are more natural from a higher loop perspective, such as the $\beta$ function of ref.~\cite{loop5pt} used in \eqn{rep52}. In the absence of more constructive methods, mining multi-loop data may be the most efficient way to currently identify symmetric representations at tree-level.   The ultimate hope, however, would be to find a constructive solution of these types of functional relations at any loop-level without relying on the introduction of a spanning ans\"atze.  It seems that finding tree-level representations may be an ideal proving ground for such techniques, where in some sense all the necessary data is available in conveniently packaged forms (color-ordered partial amplitudes and propagators), yet where the challenges share many quantitative features with finding loop-level representations.


\section*{Acknowledgements}
We thank  Zvi Bern, Lance Dixon,  Henrik Johansson,  Daniel Freedman,  Renata Kallosh, and Radu Roiban for stimulating discussions on this work and related subjects.  We are especially indebted to Daniel Freedman for collaboration during early stages of this work.    A portion of this work was completed at the Kavli Institute for Theoretical Physics, which the authors thank warmly for its hospitality.  J.B. acknowledges support from the Alexander-von-Humboldt foundation within the Feodor-Lynen program.   J.J.M.C. gratefully acknowledges the Stanford Institute for Theoretical Physics and NSF Grant No.~0756174 for financial support.  This research was supported in part by the National Science Foundation under Grant No.~NSF PHY05-51164.

\newpage

\end{document}